\documentclass{PoS}

\title{Precision charm physics, $m_c$ and $\alpha_s$ from lattice QCD }

\ShortTitle{Charm physics}

\author{C. T. H. Davies\thanks{Speaker, email: c.davies@physics.gla.ac.uk}, E. Gregory, I. Kendall, C. McNeile\\
        Dept. of Physics and Astronomy, University of Glasgow, Glasgow, UK}

\author{G. P. Lepage\thanks{Speaker, email: g.p.lepage@cornell.edu}\\
        LEPP, Cornell University, Ithaca, NY, USA}

\author{I. Allison, R. Woloshyn\\
        TRIUMF, 4004 Wesbrook Mall, Vancouver, BC, Canada}

\author{E. Dalgic, H. Trottier\\
        Simon Fraser University, Vancouver, BC, Canada}

\author{E. Follana\\
        Departmento de Fisica Teorica, Universidad de Zaragoza, Zaragoza, Spain}

\author{R. Horgan\\
        DAMTP, CMS, University of Cambridge, Cambridge, UK}

\author{K. Hornbostel\\
        Southern Methodist University, Dallas, TX, USA}

\author{J. Shigemitsu\\
        Physics Department, The Ohio State University, Columbus, OH, USA}

\author{HPQCD collaboration\thanks{http://www.physics.gla.ac.uk/HPQCD}}

\abstract{We present an update of results from the HPQCD collaboration on charm physics using 
the Highly Improved Staggered Quark action. This includes a precise determination of $m_c$ 
using moments of current-current correlators combined with high-order continuum QCD perturbation theory. We also include an update on the determination of $\alpha_s$ from lattice QCD, preliminary results on the determination of $m_b$ and a summary plot of the status of 
the gold-plated meson spectrum. There is an appendix on tackling systematic errors in fitting using the Bayesian approach. }

\FullConference{The XXVI International Symposium on Lattice Field Theory\\
		 July 14-19 2008\\
		 Williamsburg, Virginia, USA}

\begin{document}

\section{Introduction}

Lattice QCD is now established as a precision tool for calculations of the properties of 
`gold-plated' hadrons. This enables us to test QCD at a much more stringent level than 
is possible for `QCD-inspired' models and is a necessary prerequisite to trust 
lattice QCD in more speculative calculations. It also enables us to determine accurately the 
parameters of the Standard Model related to quarks, including quark masses, the strong 
coupling constant and elements of the CKM matrix. 

Recently it has become possible to handle charm quarks accurately 
in lattice QCD using the Highly Improved Staggered Quark (HISQ) action~\cite{hisq} 
and this adds an additional new dimension to this programme. 
Since charm quarks straddle the region between light quark physics and 
heavy quark physics, special care must be taken. With the HISQ action 
charm quarks are treated in the same way as light quarks and this has a 
number of advantages, discussed below. Our recent results on 
charm-light decay constants~\cite{fdetc} show the power of this approach
with results of comparable accuracy to those from light-light decay constants. 
Further results in charm physics are described in Section 2 and then 
a new method for determining the charm quark 
mass (to 1\%) using our results. Preliminary results on the determination 
of $m_b$ using the same method along with numerical results with 
the NRQCD action for $b$ quarks are also 
given. An update on the determination of $\alpha_s$ from Wilson loops completes
section 3.  The conclusions includes a summary plot of the current 
status of the gold-plated meson spectrum from lattice QCD. 
All of these results use Bayesian methods to 
allow a unified treatment of systematic errors from unknown higher 
order terms in fitting functions. Ignoring these terms clearly underestimates 
their effect; including them enables much more robust extrapolations and 
error estimates. How to do this is described in the Appendix.  

\section{Charm physics with HISQ quarks}

Charm quarks are heavy, but not very heavy, and this makes a lattice QCD 
approach to them difficult. A heavy quark treatment effectively removes the mass 
as a dynamical scale, and thereby the discretisation errors linked to it. A light 
quark treatment allows the determination of the mass from the energy at 
zero momentum and, given an action with enough chiral symmetry, conserved currents
that do not need renormalisation. As lattices become finer and finer, the charm 
quark mass in lattice units becomes smaller and the advantages of treating 
charm quarks as light relativistic quarks become more apparent. For example, 
on the MILC superfine ensemble ($a \approx$ 0.06fm) $m_ca$ is around 0.3, 
which is clearly much less than 1. The key issue is that of discretisation 
errors. These will appear as powers of $m_ca$ and can be extrapolated away, given 
results at enough values of $a$. However, if they are large i.e. low powers of 
$m_ca$ are present, so that a large extrapolation is necessary, 
they will cause significant errors in the extrapolated result. 
For a value of $m_ca = 0.4$, tree-level errors at $(m_ca)^2$ could cause a 
20\% error and $\alpha_s(m_ca)^2$ a 6\% error. It is therefore very important 
to use a highly improved action, in which these terms are removed, for an 
accurate continuum extrapolation. 

The HISQ action has been developed with exactly these issues in mind. It includes a 
further application of the `Fat7' type gluon-link smearing beyond that 
used in the improved staggered 
(asqtad) action. As for that action it includes the improvement 
of the derivative using the Naik 
3-link term which removes tree-level $a^2$ errors. 
For charm quarks the coefficient of the Naik term is further 
tuned nonperturbatively in HISQ to give 1 as the value for 
the `speed of light'. This removes the leading (in terms of the 
velocity of the charm quark inside the meson) terms at 
$\alpha_s(m_ca)^2$ and $(m_ca)^4$. 
Further details are given in~\cite{hisq}. 

\begin{figure}
\begin{center}
\includegraphics[width=8.5cm,angle=-90]{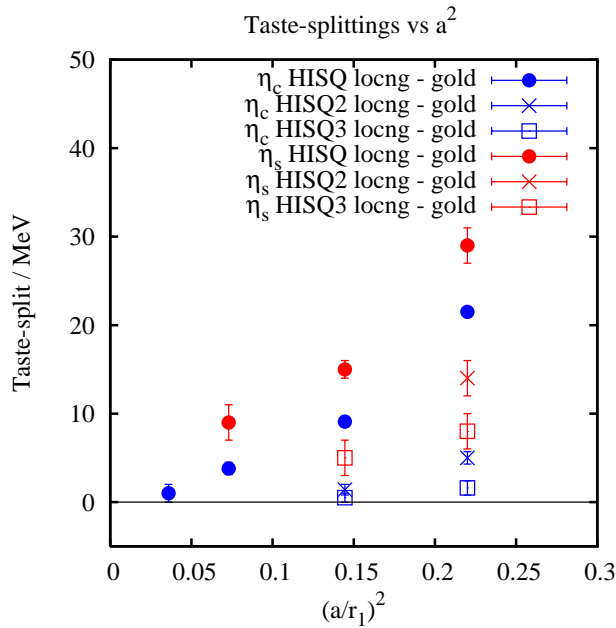}
\end{center}
\caption{The difference in mass between the goldstone pseudoscalar
and the next lightest taste, which is the local nongoldstone generated
from the local temporal axial current
for $s\overline{s}$ ($\eta_s$) and $c\overline{c}$ ($\eta_c$) and a variety 
of staggered actions. HISQ is the Highly Improved Staggered Quark action 
and HISQ2 and HISQ3 include one or two further levels of Fat7 smearing and 
associated reunitarisation and adjustment of the Lepage discretisation correction. }
\label{fig:tastes}
\end{figure}

The HISQ action is a staggered action and so has multiple 
`tastes' of quarks and mesons made from them, whose differences 
vanish as $a \rightarrow 0$. One test of 
potential discretisation errors from `taste-changing' effects 
is to study the mass splittings between different tastes of 
pseudoscalar meson. These effects show up there and are very 
small elsewhere in the spectrum. Figure~\ref{fig:tastes} shows the splitting 
between the `goldstone' $\eta_c$ and $\eta_s$ and the next lightest meson (which 
is made with the local temporal axial current) as a function 
of lattice spacing. The taste-splitting is smaller for charmonium 
than for strange-onium because the mass is heavier and both are 
clearly falling rapidly with $a^2$. We also show the taste-splittings 
with actions called `HISQ2' and `HISQ3' in which further 
applications of the Fat7 smearing are made. It is clear that the 
taste-splittings can be reduced to negligible levels, even on 
quite coarse lattices, with this procedure. 

The HISQ action has enough chiral 
symmetry for conserved vector and partially conserved axial 
vector currents so that matrix elements can be calculated on the lattice with no 
need for renormalisation. This removes a major source of systematic error. 
The HISQ action is also numerically very fast and so accurate 
results can be obtained quickly using variance reduction techniques 
such as the `random wall' previously used for light quarks~\cite{milc}. 
We fit correlators using a Bayesian multi-exponential approach~\cite{fdetc} 
taking account of oscillating states for the charm-light case (they are 
not present for charmonium). The appendix describes the Bayesian approach 
in general terms. The form for the fitting function including normal (even) 
and oscillating (odd) states is 
\begin{equation}
G(t) = \sum_n c_n (-1)^{nt}(e^{-M_nt} +  e^{-M_n(T-t)})
\label{eq:expfit}
\end{equation}
We generally take our ground state energies and amplitudes from fits that include 
4 normal (and 4 oscillating if appropriate) exponentials. By this stage the 
fit results and errors are stable as a function of the number of exponentials, 
so it is irrelevant if more are added. 

We have tuned the charm quark mass in the HISQ action using the 
goldstone $\eta_c$ mass on very coarse, coarse and fine MILC ensembles 
including 2+1 flavors of sea quarks with the asqtad action and various 
light quark masses. The scale is determined using MILC $r_1/a$ values~\cite{milcr1} and 
taking $r_1 = 0.321$ fm~\cite{alan}.
We will also include here in some plots 
preliminary results from the MILC superfine (0.0036/0.018) ensemble.
By combining HISQ charm quark propagators with light valence HISQ 
quarks we obtained accurate results for the mass of the $D$ and 
$D_s$ mesons in~\cite{fdetc}. These masses have no free parameters 
because $m_c$ was already tuned from the $\eta_c$ and they provide a 
very strong test that charmonium and charm-light physics is handled 
simultaneously correctly in lattice QCD.  Thinking of charm quarks 
as heavy quarks makes this a non-trivial test, not met by any 
continuum model of these systems. 

Our results for the $D$ and $D_s$ decay constants, made in 2007, have 
caused a lot of interest this year following new results from CLEO-c~\cite{cleo-c, cleo-c-ichep}. 
We calculate the decay constant from the matrix element of the 
partially conserved axial current in the standard way used for 
$f_{\pi}$ and $f_K$. Discretisation errors are somewhat larger for 
$D/D_s$ than for $\pi/K$ but clearly still under good control and 
extrapolation to the continuum limit from 3 values of the lattice 
spacing gives a 2\% accurate result. Again we use Bayesian methods 
for a combined chiral and continuum extrapolation 
as described in the appendix. For $D$ mesons the chiral 
extrapolation is significant; for $D_s$ mesons there is very little 
physical sea quark mass dependence, as expected. 
The decay constant is the quantity calculable in QCD that parameterises 
the annihilation rate of charged pseudoscalars to W bosons and thereby 
to leptons. 
Experimentalists can then also determine 
decay constants from the decay rate to leptons given a value for the appropriate 
CKM element from elsewhere. CLEO-c 
do this for the $D$ and $D_s$ mesons 
(measuring the rate either to $\mu \nu$ or to $\tau \nu$ and applying a small 
correction for electromagnetic effects) 
using $V_{cs} = V_{ud}$ and $V_{cd}=V_{us}$.  
Their new value for $f_D$ agrees well with our predicted result, their 
value for $f_{D_s}$ disagrees at the level of $3\sigma$ where 
$\sigma$ is dominated by the experimental error coming from 
limited statistics. 

\begin{figure}
\begin{center}
\includegraphics[width=8.5cm,angle=-90]{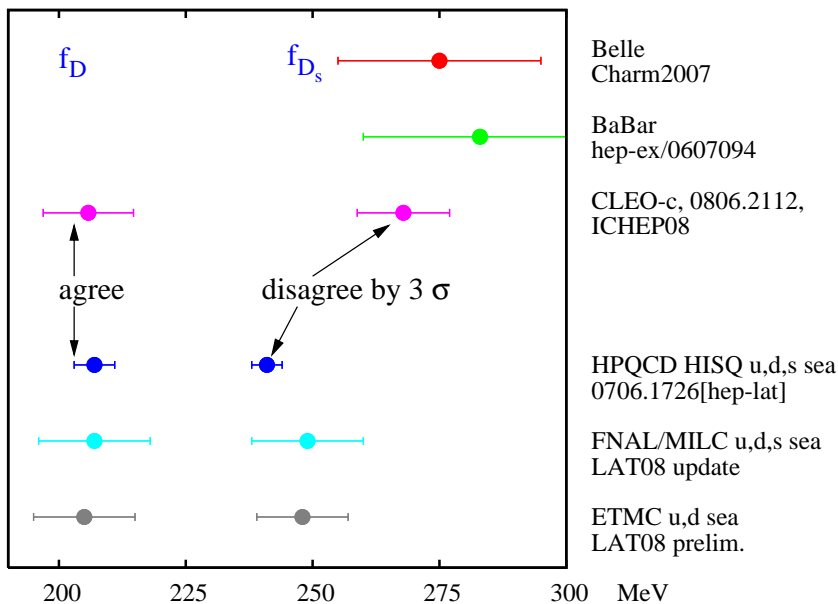}
\end{center}
\caption{A comparison of lattice results for the $D$ and $D_s$ decay constant and 
experimental results obtained from the leptonic decay rate using CKM elements $V_{cs}$ and 
$V_{cd}$ from elsewhere. The FNAL/MILC results have been updated this year by~\cite{mackenzie} 
and the ETMC results are new and described in~\cite{tarantino}. They include only 2 flavours 
of sea quarks, so are not directly comparable to the results above. There is agreement between 
lattice and experiment for $f_D$, but not for $f_{D_s}$. }
\label{fig:fdcomp}
\end{figure}

Figure~\ref{fig:fdcomp} shows a compilation of lattice results 
for the decay constants, including an update from FNAL/MILC~\cite{mackenzie} 
and a new value from $n_f=2$ calculations by ETMC~\cite{tarantino}.
Also included are the new results from CLEO-c this year~\cite{cleo-c,cleo-c-ichep}, and older 
results on $f_{D_s}$ from BaBar~\cite{babar} and Belle~\cite{belle}. There is clearly tension 
between the lattice results for $f_{D_s}$ and the experimental ones. 
This is the first time that lattice QCD has disagreed with experiment 
on a gold-plated quantity, and over 15 such quantities have now been calculated 
accurately, for examples see figure~\ref{fig:gold}. 
It could be a harbinger of new physics~\cite{kronfeld}. 
At the very least it requires everyone to check their results 
and errors thoroughly. 

\begin{figure}
\begin{center}
\includegraphics[width=8.5cm,angle=-90]{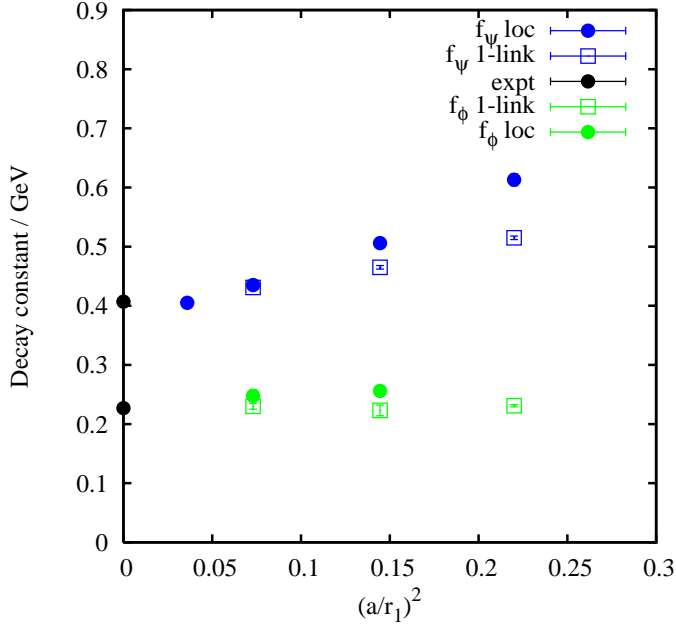}
\end{center}
\caption{Results for the decay constant for the vector $c\overline{c}$ ($J/\psi$) and 
$s\overline{s}$ ($\phi$) using currents of different taste. The $\psi$ case 
is renormalised `nonperturbatively' using results from the comparison of 
the correlators to continuum perturbation theory. The $\phi$ case is 
renormalised using one-loop lattice QCD perturbation theory (no error is 
included for unknown higher orders in the renormalisation in this plot because 
such an error is correlated between the points at different lattice spacing). 
No dependence on the light sea quark mass is seen in either of these quantities, 
as expected.}
\label{fig:lept}
\end{figure}

To this end, as well as running at a fourth superfine lattice 
spacing, we have been examining other gold-plated quantities in 
charm physics calculable using the HISQ action. 
Figure~\ref{fig:lept} shows a calculation of the decay constant 
of two vector mesons, the $\psi$ and the $\phi$. The decay constant 
can be extracted from the experimental leptonic width using
\begin{equation}
\Gamma = \frac{4\pi}{3}\alpha_{QED}^2e_Q^2\frac{f_V^2}{m_V}
\end{equation}
We compare results using different tastes of vector meson, `local' and 
(taste-singlet) `1-link'. Neither is the conserved current and 
so the lattice results had to be renormalised to compare to experiment. 
For the charmonium case this was done `nonperturbatively' (i.e. using 
only continuum perturbation theory and not lattice perturbation theory) 
making use of current-current correlators as described in the next section. 
For the $\phi$ the renormalisation was done using 1-loop lattice perturbation 
theory, which apparently has small coefficients in the HISQ case. 
Agreement with experiment is clear, work is ongoing on the error 
budget. 

\section{Determination of $m_c$, $m_b$ and $\alpha_s$}

The accurate determination of quark masses is important for 
several continuum QCD calculations. This is particularly
true of the $b$ and $c$ quark masses whose uncertainty 
strongly affects the determination, for example, of 
$V_{ub}$ from inclusive $B \rightarrow \pi$ decays. 
The limitation on this determination from standard lattice QCD methods is 
often the matching from the lattice bare mass to a continuum
scheme such as $\overline{MS}$. If lattice perturbation theory 
is used, a 2-loop determination of the matching factor must 
be done and this is hard. Nevertheless encouraging 
results can be obtained in the case of the charm quark 
using the HISQ action~\cite{allison}. 

\begin{figure}
\begin{center}
\includegraphics[width=8.5cm]{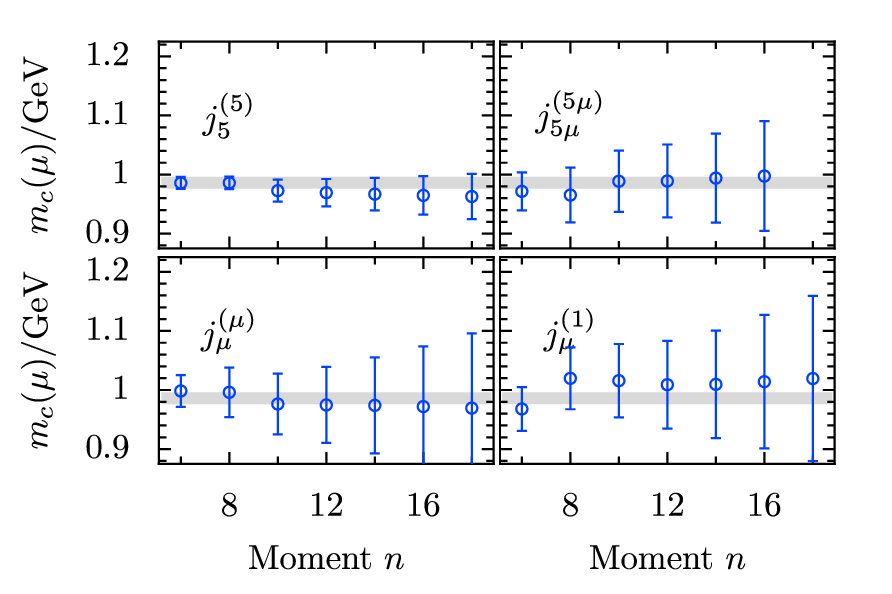}
\end{center}
\caption{$m_c(\mu)$ for $\mu$ = 3 GeV and $n_f$ = 4 flavours, from 
different moments of correlators built from four different lattice operators.
Top left is the local (goldstone) pseudoscalar, top right the local temporal axial vector and 
bottom left and right are the `1-link' vector and local vector 
respectively. The grey band is 0.986(10) GeV, which comes from 
the first two moments of the local pseudoscalar. 
}
\label{fig:mc}
\end{figure}

Here we describe a new method which takes a very different 
approach. It requires a comparison of the continuum extrapolation 
of zero-momentum lattice charmonium correlators to high order continuum 
QCD perturbation theory, and the work was done in collaboration 
with Chetyrkin, K\"{u}hn, Steinhauser and Sturm who performed 
the continuum calculations~\cite{mcjj}. The comparison is done through 
the $n$th `time-moments' of the correlators (defined below) which can be related to 
$n$th derivatives with respect to $q_0$, evaluated at $q^2=0$, 
of the polarisation function of an external current 
coupled to a heavy quark loop. This latter quantity is calculable 
in continuum perturbation theory, provided that $n$ is not too large. Our results are 
most accurate for the goldstone 
pseudoscalar ($\eta_c$) correlator. We can simply 
multiply by the square of the bare charm mass to define an 
ultra-violet finite unrenormalised (because of the PCAC relation) 
current-current correlator:
\begin{equation}
G(t) \equiv a^6 \sum_{\vec{x}}(am_{0,c})^2 <0|j_5(\vec{x},t)j_5(0,0)|0>
\end{equation}
and calculate time-moments as: 
\begin{equation}
G_n = \sum_t (t/a)^n G(t).
\end{equation}
Here $t$ goes from -$T/2$ to +$T/2$ on the lattice. 
The comparison to the continuum then becomes 
\begin{equation}
G_n(a=0) = \frac{g_n(\alpha_{\overline{MS}}(\mu), \mu/m_c)}{(am_c(\mu))^{n-4}}.
\label{eq:gncont}
\end{equation}
where $g_n$ is known through third-order in $\alpha_s$ for low values of 
$n$ (4, 6 and 8). The approach to the continuum limit is improved 
if tree level discretisation errors are removed. This is done by dividing by the 
tree level result on both sides of eq~\ref{eq:gncont}. The tree level 
lattice result is simply obtained by calculating in the free case. Tuning 
errors in $m_c$ and scale setting errors are also ameliorated by multiplying 
by factors of the lattice $\eta_c$ mass. Then the continuum extrapolation is 
actually done for
\begin{equation}
R_n \equiv \frac{am_{\eta_c}}{2am^{(0)}_{pole,c}}(\frac{G_n}{G_n^{(0)}})^{1/(n-4)}
\label{eq:Rn}
\end{equation} 
for $n \geq 6$. For $n=4$ there are no factors of $m_c$ in eq.~\ref{eq:gncont}, so we cannot 
use this to obtain $m_c$, but $\alpha_s$ can be determined as described below.
Extrapolation of the reduced time-moments, $R_n$, to the continuum limit 
again uses Bayesian methods (see appendix): 
\begin{equation}
R_n(a) = R_n(0)(1 + c_1\alpha_s(am_c)^2 + c_2\alpha_s(am_c)^4 + c_3\alpha_s(am_c)^6 + \cdots
\label{eq:extrap}
\end{equation}
We also include dependence on the sea quark masses which has negligible effect and is omitted 
here for clarity. 
Having 4 values 
of the lattice spacing from 0.06fm (MILC superfine) to 0.15fm 
(MILC very coarse) enables a very accurate continuum result 
to be obtained. The ratio of $m_c$ to the experimental $\eta_c$ mass 
is determined for different values of $n$ using: 
\begin{equation}
R_n = \frac{r_n(\alpha_{\overline{MS}}, \mu/m_c)}{2m_c(\mu)/m_{\eta_c}}
\end{equation} 
where $r_n$ is $g_n$ from eq.~\ref{eq:gncont} divided by the continuum tree 
level result. A 1\% final error is obtained by averaging results 
from $n=6$ and $n=8$. The error is dominated by uncertainties in the scale 
setting and in the perturbation theory. 
Further details are given in~\cite{mcjj}, which includes 
also determinations of $m_c$ from temporal axial and vector currents 
and a full error budget. Figure~\ref{fig:mc}, from~\cite{mcjj}, summarises 
the results. Our final number is $m_c(3 {\rm GeV}) = 0.986(10) {\rm GeV}$.
This method has been previously applied~\cite{kuehn} to determine $m_c$ 
to 1\% using experimental data on $R(e^+e^- \rightarrow {\rm hadrons})$. 
The continuum result is $m_c(3 {\rm GeV}) = 0.986(13) {\rm GeV}$. 
The fact that the lattice and continuum determinations 
of $m_c$ agree at this level of precision is a strong 
statement about how well we can handle charm quarks in 
lattice QCD using the HISQ action. 

Another application of these techniques is for the `non-perturbative'
determination of renormalisation factors for the cases where 
a nonconserved current is used in the correlator. Different moments 
have the same $Z$ factors, but different powers of $m_c$ appear. 
It is therefore possible, by taking ratios of adjacent moments 
to appropriate powers, to isolate $Z$. These factors were used 
in the calculation of the leptonic width of the $\psi$ above. 

Since the 4th moment has no powers of $m_c$ it is possible to 
relate its continuum extrapolation directly 
to a perturbative expression and use this to determine $\alpha_s$. 
We can do this also, but less accurately, for the ratios 
of powers of the 6th and 8th moments chosen to cancel powers 
of $m_c$. The value we obtain in the $\overline{MS}$ scheme 
is $\alpha_s(M_Z) = $ 0.1174(12), in good agreement 
with other determinations. More details are given in~\cite{mcjj}.

\begin{figure}
\begin{center}
\includegraphics[width=8.5cm]{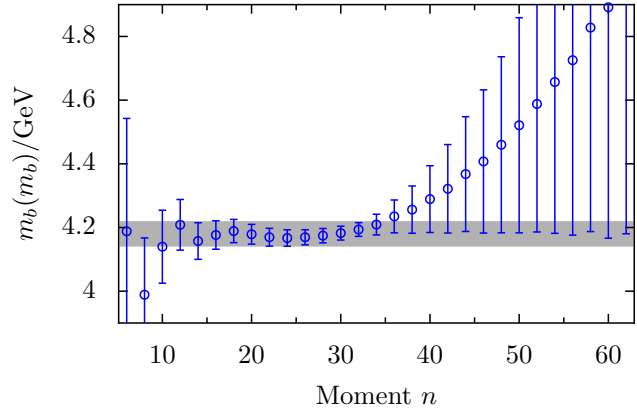}
\end{center}
\caption{Results for $m_b$ is the $\overline{MS}$ scheme at its own 
scale obtained from moments of vector ($\Upsilon$) current-current correlators 
in lattice NRQCD matched to continuum perturbation theory as described in the 
text. The shaded band is 4.18(4) GeV. At small $n$ the errors are dominated 
by relativistic and discretisation errors and at large $n$ by nonperturbative 
uncertainties.  
}
\label{fig:mb}
\end{figure}

The method of moments of current-current correlators can 
also be applied to extract $m_b$ from $\Upsilon$ and $\eta_b$ 
correlators using NRQCD $b$ quarks on the MILC configurations~\cite{kendall}. 
Now there is no conserved current in the annihilation channel and 
so ratios of moments must be taken to eliminate the factors of $Z$. 
Moments for small values of $n$ suffer from discretisation and 
relativistic corrections, but the region of validity of the perturbative 
calculation extends to larger values of $n$. A tree level analysis of 
lattice results, compared to tree level full QCD, is useful in 
understanding the systematic errors. Figure~\ref{fig:mb} shows 
preliminary results for $m_b$ from $\Upsilon$ correlators. It seems likely 
that we can achieve a 1-2\% result for $m_b$ from this method. Work is ongoing. 

\begin{figure}
\begin{center}
\includegraphics[width=8.5cm,angle=-90]{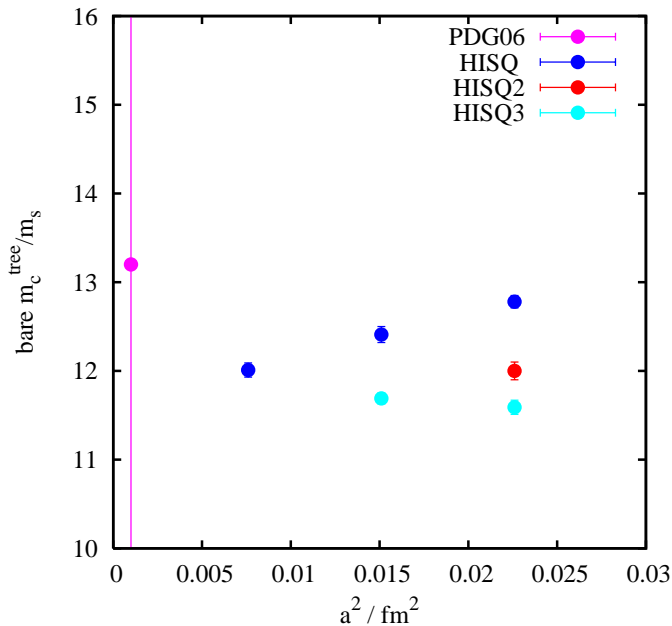}
\end{center}
\caption{The ratio of the bare lattice charm quark mass (taken as the tree level 
pole mass~\cite{hisq}) to the bare lattice strange quark mass using the HISQ 
action and variants on it. HISQ2 (double hisq) and HISQ3 (treble hisq) are described 
in the text. Matching factors to the continuum $\overline{MS}$ masses cancel in 
this ratio and allow an accurate $m_s$ to be determined. }
\label{fig:mcms}
\end{figure}

Because the HISQ action is an appropriate action for light quarks too, 
we can take advantage of cancellations of systematic errors in 
ratios of charm quantities to light ones. This allows us to leverage 
accuracy in light quark quantities from our charm results (and is therefore, 
it turns out, a very useful alternative to the more traditional 
approach of using the same action for charm and bottom and leveraging 
accuracy in bottom physics from charm~\cite{mackenzie})

Figure~\ref{fig:mcms} shows the ratio of the charm quark mass to 
the strange, obtained with HISQ, HISQ2 and HISQ3 actions 
on the MILC configurations. The matching to the $\overline{MS}$ 
scheme cancels in this ratio, when extrapolated to $a=0$. We are 
in the process of using this for an alternative accurate determination of $m_s$. 
There are clear discretisation errors in the ratio in the HISQ case 
and these are significantly reduced with the more highly smeared actions. 
Again work on this is ongoing. 

Finally, we give an update to the determination of $\alpha_s$ from 
lattice QCD using the perturbative expansion of 22 small Wilson loops or 
loop ratios~\cite{alphanew}. 
This is improved from our previous determination~\cite{alphaold} in a number 
of ways and obtains a result which is slightly more accurate and 1$\sigma$ higher 
at 0.1184(9). Our new result includes determination from 5 values of the 
lattice spacing (adding the MILC superfine and very coarse) and makes 
use of the MILC $r_1/a$ values to fix the ratios of lattice spacings 
between different ensembles more accurately. We have also fitted
the nonperturbative chiral corrections, ie. the light sea quark 
mass dependence of the logs of the Wilson loops. This is a small 
effect but certainly required by the data when using results at 
multiple sea quark masses and lattice spacing values.  The formula 
for the log of a small Wilson loop then becomes
\begin{equation}
\log(W) = \sum_{n,m} c_n \alpha_V^n(d/a) ( 1 + c_m a(2m_l+m_s) + \cdots).
\label{eq:wloop}
\end{equation}
The measured ensemble average value on the MILC ensembles is inserted 
on the left and the equation inverted to obtain $\alpha_V(d/a)$. The scale
for $\alpha$, $d/a$ is set using the BLM prescription, modified where 
necessary~\cite{kent}. 
$c_1$, $c_2$ and $c_3$ are known from the numerical evaluation of 
Feynman diagrams in lattice QCD perturbation theory, $c_4 \ldots c_{10}$ 
are constrained from our Bayesian analysis (see appendix) which fits
results for a particular $\log(W)$ from the 5 different lattice spacings 
using a common $\alpha_V$, which runs perturbatively between the different 
scales.  We are able to constrain the higher order terms somewhat more 
accurately than in our previous calculation because of having the additional 
superfine results, and these are largely responsible for the shift 
upwards of our number. The inclusion of unknown higher order terms is necessary 
for an accurate analysis, and the Bayesian approach allows us both to 
parameterise our uncertainty about them as well as allowing the data to 
constrain them where it can. 
They give sizeable corrections and without them a poor 
fit would be obtained across the multiple lattice spacings. 
The Bayesian fit also allows us to include uncertainties in the scale 
(both statistical and systematic) and analyse the effect of gluon 
condensates of various kinds. Further details are given in~\cite{alphanew}.
See also~\cite{maltman} for a somewhat different analysis of the 
perturbation theory for a subset of the Wilson loops and which gives a result in good 
agreement. 

Note the difference in methodology between the two determinations of 
$\alpha_s$ that we give here. The first (moments of current-current 
correlators) uses a quantity which is defined in the continuum 
and has lattice artefact (discretisation) errors which must be 
extrapolated away before being compared with continuum perturbation 
theory. The second (small Wilson loops) uses a quantity defined on 
the lattice and calculated in lattice perturbation theory which includes 
all lattice artefacts, order by order in $\alpha_V$. Discretisation errors 
only enter here through the presence  of such errors in the quantity 
used for scale setting in the scale for $\alpha_V$. The second method then 
has an advantage provided that lattice perturbation theory can be done 
to high enough order, and our errors reflect this. 

\section{Conclusions}

\begin{figure}
\begin{center}
\includegraphics[width=8.5cm,angle=-90]{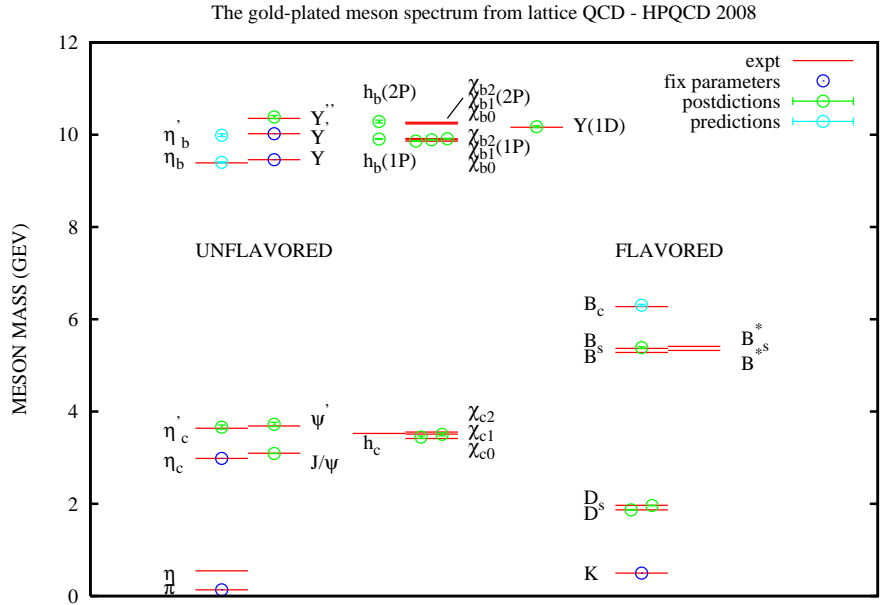}
\end{center}
\caption{The current status of the gold-plated meson spectrum. We indicate separately those 
states which are used to tune parameters (4 quark masses and the lattice spacing). We also 
show which states (the $\eta_b$ and the $B_c$) were predicted on the lattice ahead of experiment~\cite{alan, bc}. }
\label{fig:gold}
\end{figure}

Precision lattice QCD calculations continue to produce valuable 
results, building on~\cite{orig}, and we give new examples here. With the advent of the HISQ action, charm physics has 
become an excellent testing ground for lattice QCD and QCD.  
The accuracy of $b$ physics now needs to be improved to the same 
level and work on this is ongoing~\cite{eric}.
Figure~\ref{fig:gold} shows the current status of the gold-plated 
meson spectrum from HPQCD calculations.

\section*{Appendix : Constrained Fits and Error Budgets}
Constrained fitting, with Bayesian priors, is the most reliable tool for analyzing systematic errors associated with correlator fits and continuum, chiral and other extrapolations that involve a large or infinite number of increasingly unimportant terms~\cite{Lepage:2001ym}. Fitting lattice results to a chiral expansion truncated at next-to-leading order, to take a widely used example, provides little unambiguous information about the potential impact of higher-order terms in the expansion. In a constrained fit, an arbitrary number of higher-order terms can be included and their potential impact on systematic errors easily quantified.

The key to constrained fitting lies in the Bayesian priors that are included in the $\chi^2$ function that 
is minimized in the fit. Typically one is trying to fit a set of data points, 
say $Y_i\pm\sigma_{Y_i}$ at points $X_i$ for $i=1\ldots N_Y$ (for example, 
simulation results for correlators as a function of $t$ or values for $R_n$ defined in eq.~\ref{eq:Rn} 
or $\log(W)$ for different lattice spacings), to a function $y(x;c)$ that depends upon a 
set of fit parameters $c_j$ for $j=1\ldots N_c$ (for example, amplitudes and masses in the 
sum of exponentials in eq.~\ref{eq:expfit} or coefficients of the 
powers of $(am_c)^2$ in eq.~\ref{eq:extrap} or of $\alpha_V(d/a)$ or $a(2m_l+m_s)$ in eq.~\ref{eq:wloop}). 
A complication arises when in principle $N_c$ is infinite, but in practice only the first few $c_j$s 
contribute appreciably. We usually can estimate roughly how many terms are needed, 
because we have prior knowledge about the order of magnitude of the $c_j$s. 
The challenge is to incorporate systematically this prior knowledge into the 
fitting process so that the impact of unimportant or marginally important terms can be reliably assessed and quantified.

This information enters a constrained fit through the Bayesian priors for the fit parameters. In such a fit, we vary the fit parameters $c_j$ to minimize an augmented $\chi^2$ function,
\begin{equation}
	\chi^2(c) = \sum_i
	\frac{(Y_i - y(X_i;c))^2}{\sigma_{Y_i}^2} + \sum_j \delta \chi^2_{c_j} ,
\end{equation}
where there is one prior $\delta\chi^2_{c_j}$ for every fit parameter. The priors incorporate our prior knowledge about the fitting parameters. Typically one uses a Gaussian prior,
\begin{equation}
	\delta \chi_{c_j}^2 = \frac{(c_j - \overline{c}_j)^2}{\sigma_{c_j}^2},
\end{equation}
to constrain the fit parameter $c_j$ to the vicinity of $\overline{c}_j\pm\sigma_{c_j}$.

The mean $\overline{c}_j$ and width~$\sigma_{c_j}$ encode our prior knowledge about fit parameter~$c_j$. The data being fit will have a lot to say about some parameters, and almost nothing to say about others. In the former case, where a parameter is very sensitive to the data, usually the fit result for that parameter has an uncertainty that is much smaller than the width of its prior, and both the mean and standard deviation that come out of the fit are almost independent of the mean and standard deviation that went into the prior. In the other extreme, where the data is largely insensitive to a parameter, the mean and standard deviation that come out of the fit are approximately equal to the mean and standard deviation put into the prior. In the first case, the data gives us new information beyond what we put into the priors (that is, beyond what we knew before doing the fit); in the second case, the data adds nothing to our knowledge about the parameter. Parameters of the second kind, which have little relevance to the data, are good examples of \emph{nuisance} parameters in statistics.

Bayesian analysis provides a logically coherent framework for this kind of analysis~\cite{Lepage:2001ym}. Its power lies in the fact that we can include an arbitrary number of parameters without destabilizing the fit. This is as it should be. If a set of parameters is irrelevant because their corresponding terms in $y(x;c)$ are negligible, we should be free to include them or not in an analysis. It should make no difference. If they are marginally important it should have a marginal impact on the fit. This is precisely the situation when proper priors are included.

As one adds $c_j$s in the fitting function, the quality of the fit should initially improve 
(that is $\chi^2$ decrease), and usually the errors on fit results increase. Eventually, 
however, neither the fit nor the errors change when further parameters are added. 
This is the point at which the parameters become insensitive to the data 
(usually because the data is insufficiently accurate to resolve them). 
This typically occurs around the point where $\chi^2/N_Y$ falls below one. 
It is important to add terms up to the point where further terms don't change the results, 
so that systematic errors are not underestimated. 
Terms can be added beyond this point but there is little merit in this since they have no effect on fit results (means or errors).

Means and standard deviations for the fit parameters, and functions of them, 
are obtained from the minimum $\chi^2$ in the usual fashion. 
It is often useful to decompose the error~$\sigma_g$ for for some fit 
result~$g(c)$ into component parts. When errors are small (as here), 
the variance $\sigma_g^2$ is approximately linear in the variances 
that appear in the various terms in the $\chi^2$ function:
\begin{equation} \label{eq:decomposition}
	\sigma_g^2 \approx \sum_i d_{Y_i}\sigma_{Y_i}^2 + \sum_j d_{c_j} \sigma_{c_j}^2.
\end{equation}
The first sum, for example, is the contribution to the error in $g(c)$ coming from statistical errors in the data. The second sum is the contribution from uncertainties in the fit function. This sort of information is obviously very useful when thinking about improvements to an analysis\,---\,for example, in deciding how much improvement would come from better statistics for the data.

To isolate the part of the total error $\sigma_g$ that comes from, for example, the statistical uncertainty in all the $Y_i$s, the fit is rerun but with the corresponding variances in the $\chi^2$~function rescaled by a factor $f$ close to one ($f=1.1$ or 1.01, for example):
\begin{equation}
	\sigma_{Y_i}^2 \to f \sigma_{Y_i}^2
\end{equation}
for $i=1\ldots N_Y$. Then
\begin{equation}\label{eq:diffeq}
	\frac{\sigma_g(f)^2 - \sigma_g^2(f\!=\!1)}{f-1} \approx 
		\sum_{i} d_{Y_i} \sigma_{Y_i}^2.
\end{equation}
The square root of this quantity is the part of the total error that comes from statistical uncertainties in the~$Y_i$. This procedure can be repeated for each prior or group of priors in the $\chi^2$ function, thereby generating a complete error budget for the~$g(c)$. The sum of the variances obtained in this way for each part of the total error should equal~$\sigma_g^2$; if it does not, errors may not be small enough to justify the linear approximation in~\ref{eq:decomposition}~\footnote{Occasionally the difference in~\ref{eq:diffeq} comes out negative. This could be caused by instabilities in the fit, in which case changing $f$ might change the sign. It is possible, however, for one of the coefficients in~\ref{eq:decomposition} to be negative. In such cases we use the absolute value of~\ref{eq:diffeq} for the variance.}.

Most of the uncertainties in a standard lattice analysis can be pushed into the constrained fit, to 
facilitate consistent treatment of all systematic and statistical errors. 
For example, the independent variables $X_i$ might have errors:  $X_i = \overline{X}_i\pm\sigma_{X_i}$. 
The determination of the lattice spacing will typically have a systematic error 
that grows with the lattice spacing that can be taken into account this way, for example. 
We do this in the analysis of Wilson loops to derive $\alpha_V$ since no continuum limit 
is taken there but the systematic errors in the comparison of scales between ensembles is 
very important. Another example is that of chiral extrapolation using quark masses, where 
the quark masses have to be run to a common scale and will therefore have systematic 
errors associated with them. 
In these cases we treat each $X_i$ as another fit parameter, 
to be varied in the fit, and include the following priors in the $\chi^2$ function:
\begin{equation}
	\delta \chi^2_{X} = \sum_i \frac{\left(X_i-\overline{X}_i\right)^2}{\sigma_{X_i}^2}.
\end{equation}
One of the outputs from the fit might be improved values for the~$X_i$.

There are usually a variety of other parameters that introduce errors into a lattice analysis. 
For example, we take the lattice spacing from MILC data for $r_1/a$ 
and use our $\Upsilon$ determination of $r_1$~\cite{alan, kendall}. Each of these has errors. 
The $(r_1/a)_i$ for each data set and the overall $r_1$ are all treated as fit 
variables when we fit our lattice data to a function that depends on the lattice spacing, and each has a prior:
\begin{equation}
	\delta\chi^2_a = \sum_i \frac{\left((r_1/a)_i -
	 \overline{(r_1/a)}_i\right)^2}{\sigma_{(r_1/a)_i}^2}
	+ \frac{(r_1-\overline{r}_1)^2}{\sigma_{r_1}^2}.
\end{equation}
Here $\overline{(r_1/a)}_i$ and $\sigma_{(r_1/a)_i}$ are taken from MILC results, while $\overline{r}_1\pm\sigma_{r_1}=0.321$(5)\,fm. 

Any quantity that enters the analysis and has uncertainties is treated as a fit parameter, with a prior that incorporates whatever is known about that quantity. In this way all uncertainties can be analyzed simultaneously within a single constrained fit.


\begin{thebibliography}{99}
\bibitem{hisq} E. Follana {\it et al}, HPQCD collaboration, Phys. Rev. D{\bf 75}:054502 (2007) [hep-lat/0610092].
\bibitem{fdetc} E. Follana, C. T. H. Davies, G. P. Lepage and J. Shigemitsu, HPQCD collaboration, Phys. Rev. Lett. {\bf 100}:062002 (2008) arXiv:i0706.1726[hep-lat].
\bibitem{milc} C. Aubin {\it et al}, MILC collaboration, Phys. Rev. D{\bf 70}:114501 (2004), [hep-lat/0407028].
\bibitem{milcr1} C. Aubin {\it et al}, MILC collaboration, Phys. Rev. D{\bf 70}:094505 (2004), [hep-lat/0402030].
\bibitem{alan} A. Gray {\it et al}, HPQCD collaboration, Phys. Rev. D{\bf 72}:094507 (2005) [hep-lat/0507013].
\bibitem{cleo-c}
K. Ecklund {\it et al}, CLEO-c, Phys. Rev. Lett{\bf 100}:161801 (2008) arXiv:0712.1175[hep-ex]; B. Eisenstein {\it et al}, CLEO-c, arXiv:0806.2112[hep-ex];
\bibitem{cleo-c-ichep}
L. Zhang, CLEO-c, this Proceedings, arXiv:0810.2328[hep-ex]. 
\bibitem{mackenzie} P. Mackenzie {\it et al}, Fermilab lattice/MILC collaborations, these Proceedings.
\bibitem{tarantino} C. Tarantino {\it et al}, ETMC collaboration, these Proceedings arXiv:0810.3145[hep-lat].
\bibitem{babar}
B. Aubert {\it et al}, BaBar collaboration, Phys. Rev. Lett.{\bf 98}:141801 (2007) [hep-ex/0607094].
\bibitem{belle}
L. Widhalm, Belle collaboration, Proceedings of Charm 2007, arXiv:0710.0420. 
\bibitem{kronfeld} A. Kronfeld, these Proceedings.
\bibitem{allison} I. Allison {\it et al}, HPQCD collaboration, these Proceedings arXiv:0810.0285[hep-lat].
\bibitem{mcjj} I. Allison {\it et al}, HPQCD collaboration + K. G. Chetyrkin, J. H. K\"{u}hn, M. Steinhauser and C. Sturm, arXiv:0805.2999[hep-lat], Phys. Rev. D (in press). 
\bibitem{kuehn} J. H. K\"{u}hn, M. Steinhauser and C. Sturm, Nucl Phys. B{\bf 778} 192 (2007) [hep-ph/0702103].
\bibitem{kendall} I. Kendall {\it et al}, HPQCD collaboration, these Proceedings. 
\bibitem{alphanew} C. T. H. Davies, I. Kendall, G. P. Lepage, C. McNeile, J. Shigemitsu, H. Trottier, HPQCD collaboration, arXiv:0807.1687[hep-lat].
\bibitem{alphaold} Q. Mason {\it et al}, HPQCD collaboration, Phys. Rev. Lett. {\bf 95}:052002 (2005) [hep-lat/0503005]. 
\bibitem{kent} K. Hornbostel, G. P. Lepage and C. Morningstar, Phys. Rev. D{\bf 67}:034023 (2003) [hep-ph/0208224].
\bibitem{maltman} K. Maltman {\it et al}, these Proceedings; K. Maltman, D. Leinweber, P. Moran and A. Sternbeck, arXiv:0807.2020[hep-lat].
\bibitem{orig} C. T. H. Davies {\it et al}, Fermilab Lattice/HPQCD/MILC collaborations, Phys. Rev. Lett. {\bf 92}:022001 (2004) [hep-lat/0304004]. 
\bibitem{eric} E. Gregory {\it et al}, HPQCD collaboration, these Proceedings arXiv:0810.1845[hep-lat]; S. Meinel {\it et al}, these Proceedings arXiv:0810.0921[hep-lat]. 
\bibitem{Lepage:2001ym} G. P. Lepage {\it et al}, Nucl. Phys. Proc. Suppl. {\bf 106}, 12 (2002) [hep-lat/0110175].
\bibitem{bc}
I. F. Allison {\it et al}, HPQCD/Fermilab Lattice collaborations, Phys. Rev. Lett{\bf 94}:172001 (2005) [hep-lat/0411027].

\end{thebibliography}
\end{document}